\documentclass[useAMS,usenatbib]{mn2e}
\usepackage{epsfig}
\usepackage{graphicx}
\usepackage{times}
\usepackage{color}
\def\sun{\hbox{$\odot$}}
\def\cm3{cm$^{-3}$}
\def\kms{km~s$^{-1}$}

\def\msun{M$_{\odot}$}

\def\beq{\begin{equation}}
\def\eeq{\end{equation}}

\def\lesssim{\mathrel{\hbox{\rlap{\hbox{\lower4pt\hbox{$\sim$}}}\hbox{$<$}}}}
\def\gtrsim{\mathrel{\hbox{\rlap{\hbox{\lower4pt\hbox{$\sim$}}}\hbox{$>$}}}}

\def\lesssim{\mathrel{\hbox{\rlap{\hbox{\lower4pt\hbox{$\sim$}}}\hbox{$<$}}}}
\def\gtrsim{\mathrel{\hbox{\rlap{\hbox{\lower4pt\hbox{$\sim$}}}\hbox{$>$}}}}

\def\one{{\,\sc i}}
\def\two{{\,\sc ii}}
\def\three{{\,\sc iii}}
\def\four{{\,\sc iv}}

\def\cmfgen{{\sc cmfgen}}

\newcommand{\iso}[2]{\ensuremath{^{#1}\rm{#2}}}

\def\pasp{PASP}
\def\apj{ApJ}

\def\apjl{ApJL}
\def\aap{A\&A}
\def\araa{ARA\&A}

\def\mnras{MNRAS}
\def\nat{Nature}

\def\cofl{[Co\three]}
\def\nifs{\iso{56}Ni}
\def\cofs{\iso{56}Co}
\def\fefs{\iso{56}Fe}

\title[ \hbox{\rm [}Co\three\hbox{\rm ]} vs. Na\one\,D in SN Ia spectra]{[Co\three] versus Na\one\,D
in type Ia supernova spectra}

\author[Luc Dessart, D.J. Hillier, St\'ephane Blondin, and Alexei Khokhlov]
{Luc Dessart,$^{1}$ D. John Hillier,$^{2}$ St\'ephane Blondin,$^{1}$ and Alexei Khokhlov$^{3}$ \\ \\
$^{1}$Aix Marseille Universit\'e, CNRS, LAM (Laboratoire d'Astrophysique
de Marseille) UMR 7326, 13388, Marseille, France.\\
$^{2}$: Department of Physics and Astronomy \& Pittsburgh Particle physics, Astrophysics,
and Cosmology Center (PITT PACC), University of Pittsburgh,   \\
3941 O'Hara Street, Pittsburgh, PA 15260, USA. \\
$^3$ Department of Astronomy \& Astrophysics and the Enrico Fermi Institute, The University of Chicago,
Chicago, IL 60637, USA}

\begin{document}

\date{Accepted . Received }

\pagerange{\pageref{firstpage}--\pageref{lastpage}} \pubyear{2011}

\maketitle

\label{firstpage}

\begin{abstract}
   The high metal content and fast expansion of supernova (SN) Ia ejecta leads to considerable
line overlap in their optical spectra. Uncertainties in composition and ionization further complicate
the process of line identification. In this paper, we focus on the 5900\,\AA\ emission feature seen in
SN Ia spectra after bolometric maximum, a line which in the last two decades has been associated
with [Co\three]\,5888\,\AA\ or Na\one\,D. Using non-LTE time-dependent radiative-transfer calculations
based on Chandrasekhar-mass delayed-detonation models, we find that Na\one\,D line emission
is extremely weak  at all post-maximum epochs. Instead, we predict the presence of [Co\three]\,5888\,\AA\ after maximum in
all our SN Ia models, which cover a range from 0.12 to 0.87\,\msun\ of \nifs. We also find that the 
[Co\three]\,5888\,\AA\ forbidden line is present within days of bolometric maximum, and strengthens 
steadily for weeks thereafter. Both predictions are confirmed by observations.
Rather than trivial taxonomy, these findings confirm that it is necessary to include forbidden-line transitions
in radiative-transfer simulations of SNe Ia, both to obtain the correct ejecta cooling rate and to
match observed optical spectra.
\end{abstract}


\section{Introduction}

Radiative-transfer modeling of supernova (SN) Ia is a challenging enterprise
(\citealt{pinto_eastman_93,hoeflich_95,BHN96_non-lte,pinto_eastman_00a,pinto_eastman_00b,
hoeflich_03,kasen_etal_06,jack_etal_11,dessart_etal_13xx}, hereafter D13).
For a start, the initial ejecta conditions for such simulations
are uncertain. The possibility of both single- and double-degenerate progenitor systems suggests that
the ejecta mass likely varies amongst SN Ia ejecta (e.g.,
\citealt{pinto_eastman_00a,sim_etal_10,kerkwijk_etal_10}).
The ejecta composition that results from the combustion of a C/O white dwarf is not known with confidence
because the explosion scenario, besides being non unique, involves hard-to-model combustion
physics \citep{HN00_Ia_rev}.
As a result, for each scenario, simulations of the explosion, whether 1-D or multi-D,
are parametrized rather than modeled consistently from first principles
\citep{K91a,gamezo_etal_05,roepke_hillebrandt_05}.
The progenitor is likely to depart from a simple  hydrostatic configuration for many possible reasons including
fast rotation \citep{yoon_langer_04}, the conditions produced by the smoldering phase and ignition
(see, e.g., \citealt{woosley_etal_04}), or from the complex dynamical evolution in a merger
event \citep{pakmor_etal_11,pakmor_etal_12}.

Even if the ejecta properties were accurately known, the modeling of the radiation would remain
challenging because of the prevalence of line opacity, the strong
influence of scattering resulting from the very low gas densities in the fast expanding low mass
ejecta, the importance of non-LTE and time-dependent effects, and non-thermal processes.
Perhaps even more important are the numerous
processes that take place between the various constituents of the gas (electrons and ions),
involving thousands of atomic levels, and which control its thermodynamic state.
Historically, this complexity has generally been interpreted in terms of
an ``opacity" problem (see, e.g., \citealt{hoeflich_etal_93,pinto_eastman_00b,kasen_etal_08}).
In \citet{dessart_etal_13xx}, we demonstrate that the temperature and ionization distribution,
which are difficult to determine accurately, are also important (not surprisingly) 
because the thermodynamic state of the gas determines which ions provide the opacity. 
We found, for example, that forbidden
lines are crucial, as early as bolometric maximum, in controling the cooling of
the ejecta. Paradoxically, these lines have low oscillator strengths, and thus tend to be
ignored in simulations prior to $\lesssim$\,100\,d after explosion.
Finally, inaccuracies in the atomic data, or the lack of atomic data altogether,
introduce a major source of uncertainty in any radiative transfer modeling of SNe Ia.

A central problem with SNe in general is that lines are strongly Doppler broadened by
the fast expansion of the ejecta. The large metal mass fraction in SNe Ia leads to the presence
of forests of lines which overlap, typically preventing the identification of a ``clean"
line anywhere in their spectra. The emerging radiation is in addition strongly influenced by a few strong
lines, such as Si\two\,6355\,\AA\ or the Ca\two\ triplet, giving the wrong impression
that the spectrum is analogous to a blackbody influenced by a few spectral features
of large optical depth.
This situation is particularly problematic when SNe Ia evolve past maximum because at
such times, the optical depth clearly drops, the ejecta turns nebular, but strong lines are still
present. There are no regions with negligible flux, even though
the continuum optical depth is well below unity even at bolometric maximum
\citep{hoeflich_etal_93,pinto_eastman_00b,hillier_etal_13}.

After bolometric maximum, the peaks and valleys of SN Ia spectra are a complex blend
of numerous lines, some thick, others thin, each interacting with hundreds of other lines
either locally (within a Sobolev length) or non-locally (because redshifted into resonance
with a redder line). This conspires to produce confusion about spectrum formation in SNe Ia.
The concept of a ``photosphere" is routinely used but the notion of a sharp boundary from where
radiation would escape does not hold for SNe Ia.
\citet{branch_etal_08} propose that  most/all lines at nebular times are
permitted (generally resonance) transitions, while numerous papers emphasize the near
exclusive presence of forbidden-line transitions \citep{kuchner_etal_94,mazzali_etal_08}.

To give additional evidence for the importance of forbidden lines in SNe Ia \citep{dessart_etal_13xx},
we focus here on the 5900\,\AA\ feature observed in post-maximum SN Ia spectra.
In recent years, this feature has been associated with Na\one\,D, although this association remains
suspicious and puzzling. Models sometimes provide a very good
fit \citep[Fig.~3]{mazzali_etal_08}, a poor fit \citep{branch_etal_08}, or predict no feature
at that location  \citep{kasen_etal_09, blondin_etal_11,tanaka_etal_11}.
In the  next section, we give an historical perspective of earlier work that
modeled or discussed the spectral feature at 5900\,\AA.
We then present results from our grid of delayed-detonation and pulsational-delayed-detonation
models covering a range of \nifs\ mass and show that this feature can be explained
as [Co\three]\,5888\,\AA\ emission, across the range from sub-luminous to standard SNe Ia
(Section~\ref{sect_res}). In other words, we find that such ejecta models naturally produce
an emission feature at 5900\,\AA, and that this emission in our models is systematically
associated with [Co\three]\,5888\,\AA.
Our conclusions and a discussion of future work are presented in Section~\ref{sect_conc}.

Far from being a banal characteristic of SN Ia spectra, the observation of the [Co\three]\,5888\,\AA\ 
line indicates that forbidden line
transitions have to be incorporated in any SN Ia model at and beyond bolometric maximum, not just
to reproduce observations, but also to compute correctly the cooling of SN Ia ejecta.

\section{Previous works}
\label{sect_hist}

   The spectral feature at 5900\,\AA\ in SNe Ia has been discussed repeatedly in the last two
decades. We can separate the studies focusing on the ``photospheric" phase (epochs until soon after
bolometric maximum) and those devoted to ``advanced" nebular phase (beyond 100\,d
after explosion when the SN exhibits an apparently pure-emission spectrum).

\citet{axelrod_80} is probably the first to study nebular-phase spectra of SNe Ia, and
he associates unambiguously the
5900\,\AA\ feature with the forbidden transition of Co\three\ at 5888\,\AA.
Later, \citet{eastman_pinto_93} present numerical developments incorporated into the
code {\sc eddington}, and apply their technique to spectrum formation in a
SN Ia ejecta at 250\,d after explosion. They propose \cofl\ as the origin of the 5900\,\AA\
feature (see their Fig.~5).
\citet{kuchner_etal_94}  use the simultaneous presence of [Co\three]\,5888\,\AA\ and
[Fe\three]\,4658\,\AA\ (strictly speaking, the 4500-5000\,\AA\ region contains lines from both Fe\two\
and Fe\three)
in SN Ia spectra to confirm the radioactive decay of \nifs\ at the origin of the SN Ia luminosity.
Indeed, they find the flux ratio of
these two lines (ignoring line overlap and some complications of the radiative transfer) can be explained
from the decay of \cofs\ to \fefs. They discard the possible association of the 5900\,\AA\ feature with Na.
\citet{mazzali_etal_97} study SN\,1991bg at both early times and late times.
At nebular times, they propose that the 5900\,\AA\ feature is primarily associated with [Co\three], and
show how this Co forbidden transition may be used to
set constraints on the original \nifs\ mass. Their nebular model reproduces
the 5900\,\AA\ feature, in both strength and width (see their Fig.~14), supporting the same
assessment made by \citet{kuchner_etal_94}.

\begin{table}
\caption{Summary of nucleosynthetic yields for the Chandrasekhar-mass delayed-detonation
models used in this work. Numbers in parenthesis correspond to powers of ten.}\label{tab_modinfo}
\begin{center}
\begin{tabular}{lc@{\hspace{1.8mm}}c@{\hspace{1.8mm}}c@{\hspace{1.8mm}}c@{\hspace{1.8mm}}c@{\hspace{1.8mm}}}
\hline\hline
\multicolumn{1}{c}{Model} & $M$(\nifs) & $M$(Ni) & $M$(Fe) & $M$(Si) & $M$(Na) \\
 & [M$_{\sun}$] & [M$_{\sun}$] & [M$_{\sun}$] & [M$_{\sun}$] & [M$_{\sun}$]  \\
\hline
DDC0           &       0.869 &      0.872 &        0.102 &       0.160 &   6.22(-6)  \\
DDC6           &       0.722 &      0.718 &        0.116 &       0.216 &   1.02(-5)  \\
DDC10          &       0.623 &      0.622 &        0.115 &       0.257 &   1.26(-5)  \\
DDC15          &       0.511 &      0.516 &        0.114 &       0.306 &   1.70(-5)  \\
DDC17          &       0.412 &      0.421 &        0.112 &       0.353 &   2.53(-5)  \\
DDC20          &       0.300 &      0.315 &        0.110 &       0.426 &   3.53(-5)  \\
DDC22          &       0.211 &      0.231 &        0.107 &       0.483 &   6.30(-5)  \\
DDC25          &       0.119 &      0.142 &     9.80(-2) &       0.485 &   1.51(-4)  \\
\hline                                                                             
PDDEL3         &       0.685 &      0.680 &        0.107 &       0.218 &   1.57(-5)  \\
\hline
\end{tabular}
\end{center}
\end{table}

\begin{figure*}
\epsfig{file=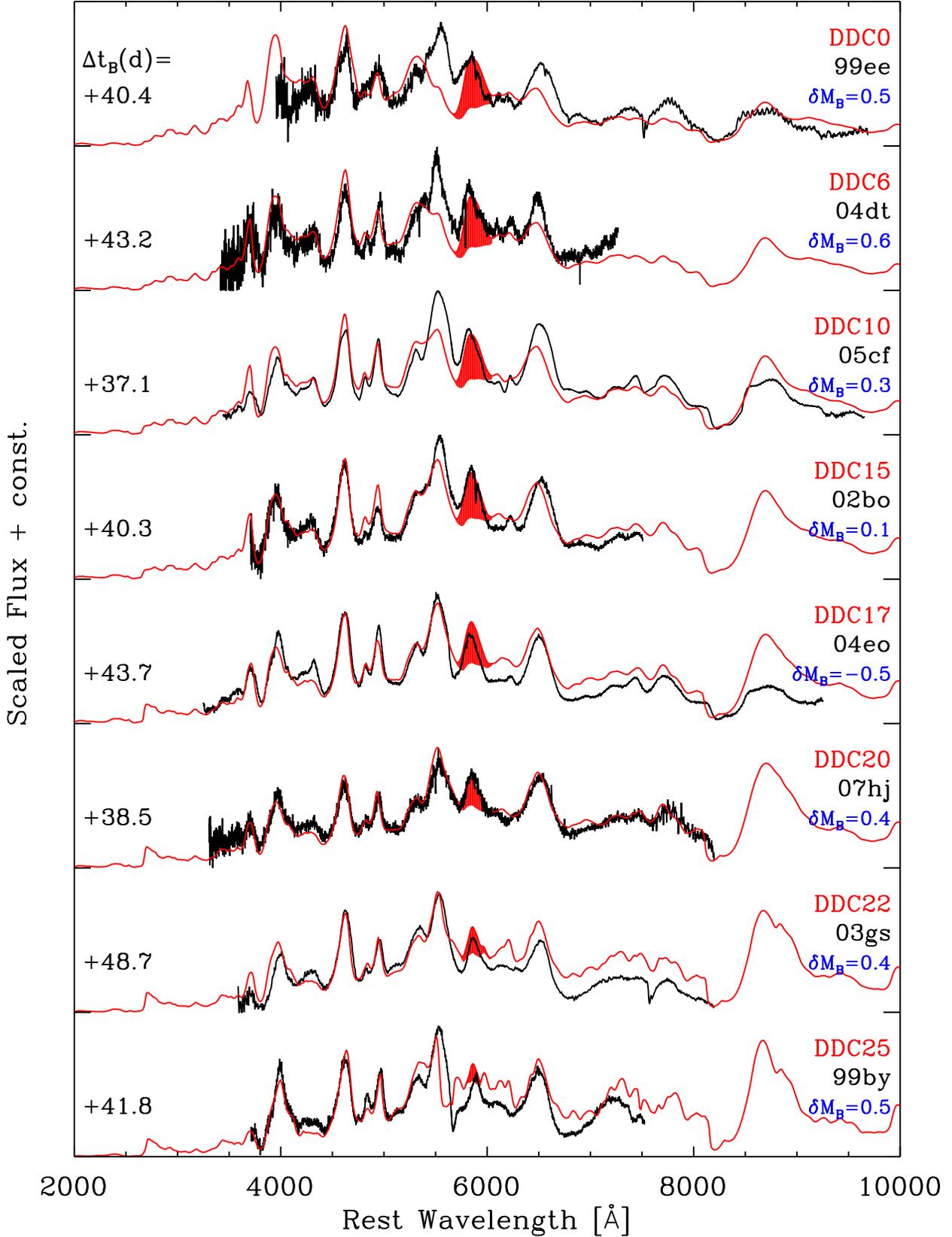,width=16.5cm}
\caption{Montage of spectra comparing our grid of delayed-detonation models (red) with observations (black)
at $\sim$\,40\,d after $B$-band maximum (see left label). 
This plot is a counterpart of Fig.~5 of  \citet{blondin_etal_13} which makes a similar comparison near bolometric maximum
(we have replaced SN\,1992A by SN\,2004eo since the former does not have data at this post-explosion epoch).
For each pair, we give the $B$-band magnitude offset at the corresponding time (blue label on the right).
The filled area colored in red corresponds to the synthetic flux associated with the [Co\three]\,5888\,\AA\ forbidden-line transition.
\label{fig_grid}
}
\end{figure*}

Disparate interpretations seem to start with the work of \citet{lentz_etal_01} who fail to reproduce the 5900\,\AA\
region of SN\,1994D at early post-peak epochs, despite strong modulations of the adopted Na abundances
in their ejecta model. They suggest that the explosion model employed is probably inadequate,
but it also seems that they do not include forbidden-line transitions in the radiative-transfer modeling.
\citet{branch_etal_08} use {\sc synow} to identify the lines present
in SNe Ia after bolometric maximum and conclude that most lines are permitted.
They also argue for a Na\one\,D association with the 5900\,\AA\ feature
(the broad absorption they predict is however associated with a broad line emission
that  is unseen), and while they mention the \cofl\ possibility they do not retain it.
\citet{mazzali_etal_08} present an analysis of SN\,2004eo spectra. They associate the
5900\,\AA\ emission feature at 250\,d after explosion with Na\one\,D.
At earlier times, they suggest the 5900\,\AA\ feature is due to Si\two\ --
it seems that no [Co\three] line is included in their modeling.
Along the same reasoning, \citet{tanaka_etal_11} tie the 5900\,\AA\ feature to Na\one\,D in SN\,2003du,
although the feature is not fitted by their model.

\citet{maurer_etal_11} perform detailed non-LTE steady-state radiative-transfer calculations
at nebular times and explicitly discuss [Co\three] line emission.
Although the strength of [Fe\three] and [Co\three] lines varies between codes,
sometimes significantly, [Co\three] remains the most plausible explanation for the 5900\,\AA\ line.

It thus seems that from being secure in the 90's, the association of the 5900\,\AA\ feature with the
[Co\three]\,5888\,\AA\ line is no longer retained, even though the Na\one\,D
association is rarely matched satisfactory by SN Ia radiative-transfer models.\footnote{Part of the confusion 
has undoubtedly arisen because forbidden line emission is generally associated with low density, and thus
the detection of a feature at 5900\,\AA\ not long after maximum would seem to preclude its identification 
with a forbidden line. A key distinction, however, is that Co is not an impurity species, and thus it is much 
more likely to be seen even though ejecta densities exceed the critical density.}
Following upon our recent study of SN Ia physics and spectrum formation (D13), we study the
origin of the 5900\,\AA\ feature in post-maximum SN Ia spectra using \cmfgen\ and our
most complete model atom (see D13 for details). For this purpose, we use
the same delayed detonation models with which we obtain good agreement to
maximum-light spectra of SNe Ia \citep{blondin_etal_13}. We also include one
pulsational-delayed-detonation model from \citet{dessart_etal_13yy}, model PDDEL3,
since it yields a satisfactory match to the SN Ia 2005cf evolution from about $-$10 to +80\,d.
In numerous ways, the match is superior to that obtained with model DDC10,
for reasons that we discuss in \citealt{dessart_etal_13yy}.

\section{Results from delayed-detonation models}
\label{sect_res}

In this paper, we use a sample of simulations from \citet{blondin_etal_13}, D13,
and \citet{dessart_etal_13yy}. All are delayed detonations (DDC sequence), but some
explode following a pulsation (PDDEL sequence; see \citealt{dessart_etal_13yy} for details).
We summarize the key nucleosynthetic yields for these models in Table~\ref{tab_modinfo}.

For the discussion of the paper, it suffices to say that these models cover a range of \nifs\ mass,
from 0.18 to 0.81\,\msun, and provide a satisfactory match to a wealth of SNe Ia at maximum
light \citep{blondin_etal_13}. A fundamental property of {\it all} our delayed
detonation models is that the sodium mass fraction below 15000\,\kms\ is on the order of 10$^{-10}$.
The total sodium mass in our models is typically on the order of 10$^{-5}$\,\msun.

In D13, we described how one must exert extreme care to account for a number of critical non-LTE
processes in order to follow the SN Ia evolution from maximum light to the nebular phase.
This not only requires a detailed account for opacity sources, especially associated with metals,
but also for critical coolants which act as primary ingredient for setting the temperature and ionization
state of the gas. So, each model we present here was evolved from 1\,d to 100\,d after explosion
with all the key processes we found to be important (D13).

It would be presumptuous to pretend that with eight delayed-detonation models, 
we could reproduce all SNe Ia that exist. Instead, our strategy is
to identify key signatures that characterize and distinguish SNe Ia, and test whether
such delayed-detonation models predict those key signatures, without any tinkering of
our hydrodynamical models (on composition, density profile, total mass, etc.).
This strategy is sound because SNe Ia constitute a highly homogeneous class of events once
we exclude the less frequent 91bg-like and 91T-like events.
If a given explosion model has any validity, it should reproduce at least a subset of SNe Ia.
We suspect this degeneracy calls for a similar degeneracy in SN Ia ejecta, and our simulations of
delayed detonations confirm this.

In Fig.~\ref{fig_grid}, we present a counterpart of the montage of  $B$-band maximum spectra presented
in \citet{blondin_etal_13}, but this time for an epoch of $\sim$\,40\,d after B-band maximum. 
At this time, SNe Ia exhibit a very different spectral morphology, with a reduced flux in the blue and 
a dominance of metal lines, in particular from iron around 5000\,\AA.
In this montage, we can clearly see that a broad feature is present around 5900\,\AA\ in all
SNe Ia selected (which are quite typical of the SN Ia population as a whole), and therefore,
that this feature is present whether we consider a luminous SN Ia like 1999ee or
a sub-luminous SN Ia like 1999by.

Interestingly, all our delayed-detonation models predict a feature near 5900\,\AA. At 40\,d past $B$-band 
maximum, the spectrum forms in the inner ejecta at velocities $\lesssim$\,10000\,\kms, although some photons 
will still experience scattering and absorption at larger velocities if they overlap with strong lines (primarily from 
intermediate mass elements) or if they lie in the UV range.
In the inner ejecta, our models have a composition dominated by cobalt and iron.
As mentioned above, delayed detonation models leave no trace of sodium below about 15000\,\kms,
and as expected we find that Na\one\,D emission is negligible in our synthetic spectra at those epochs.
To illustrate the association of [Co\three]\,5888\,\AA\
(the properties of the Co\three\ forbidden-line transitions in the 6000\,\AA\ region are given in Table~\ref{tab_co3})
with the observed feature, we recompute
the synthetic spectra but set the oscillator strength of that forbidden-line transition to zero.
The resulting synthetic spectra are indistinguishable, apart from a strong difference in the 5900\,\AA\ region.
The associated flux difference in shown as a filled area colored in red.

After the detailed discussion presented in D13, it is
not surprising that this feature is indeed primarily due to [Co\three]\,5888\,\AA.
And this explanation is more sensible from the point of view of nucleosynthesis since
the sodium mass fraction in the inner ejecta is vanishingly small in all delayed-detonation
models.

\begin{figure*}
\epsfig{file=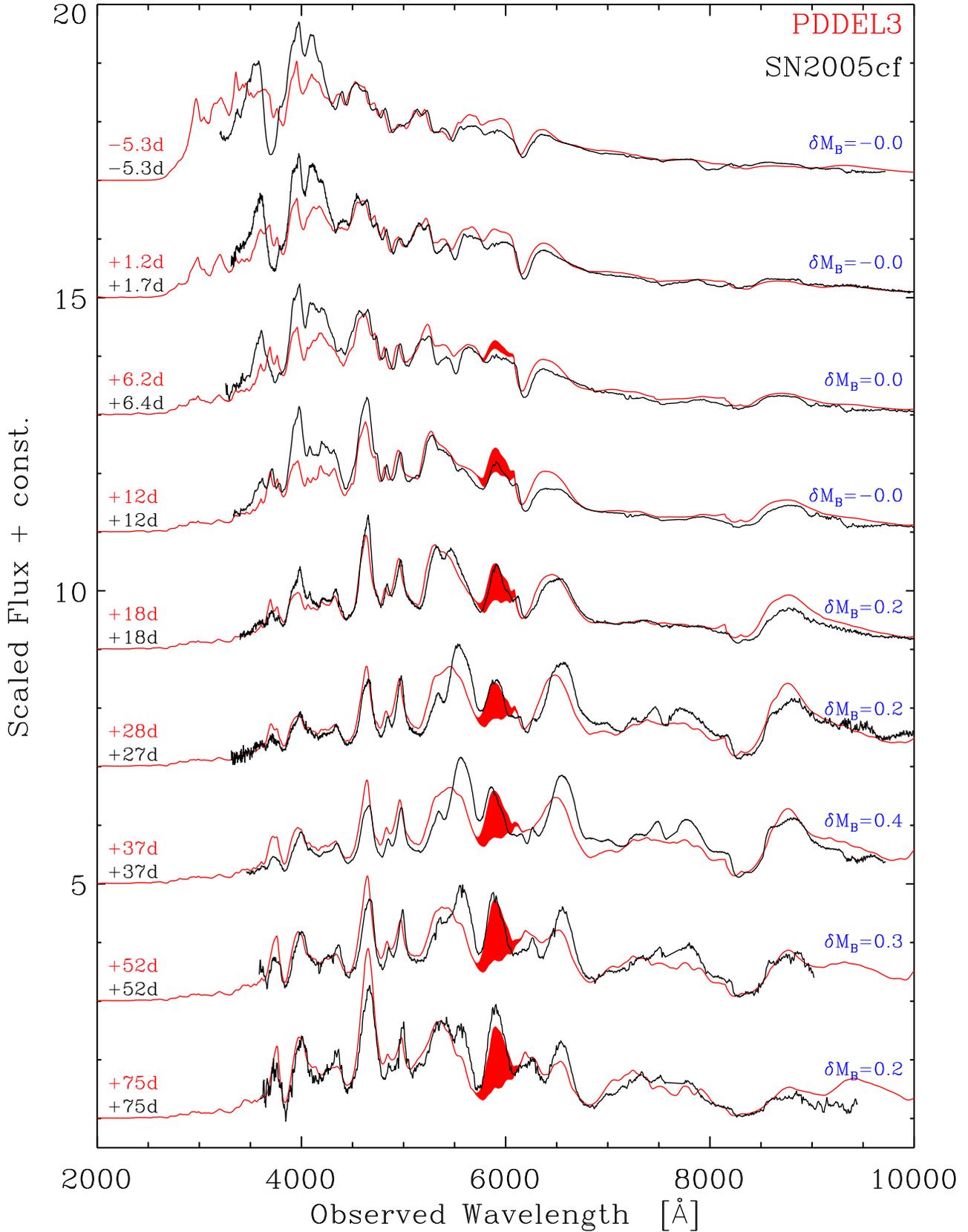,width=17cm}
\caption{
Montage of spectra comparing  the observations of SN\,2005cf (black) with model PDDEL3
(red; \citealt{dessart_etal_13yy}). The filled area colored in red corresponds to the
synthetic flux associated with the [Co\three]\,5888\,\AA\ forbidden-line transition.
Times are with respect to $B$-band maximum.
Each model spectrum is reddened, redshifted, and scaled to match the inferred distance
to SN\,2005cf. A vertical scaling is used at each epoch for better visibility. However, the label
on the right gives the $B$-band magnitude offset between model and observation
at the corresponding epoch.
\label{fig_seq}
}
\end{figure*}

\begin{table}
\caption{Summary of atomic properties for the [Co\three] lines in the 6000\,\AA\ region, including the
line at the origin of the 5900\,\AA\ feature in post-maximum SN Ia spectra. The collisional strengths of Co\three, which are not known,
are adopted from those of Ni\four\ \citep{2002CoPhC.145..311S} since it possesses a very similar term structure
\citep{NIST}. Parentheses indicate powers of ten.
}\label{tab_co3}
\begin{tabular}{l@{\hspace{1.8mm}}c@{\hspace{1.8mm}}c@{\hspace{1.8mm}}c@{\hspace{1.8mm}}c@{\hspace{1.8mm}}}
\hline\hline
  & $\lambda$ [\AA] & Transition & $f$ & A [s$^{-1}$] \\
\hline
\hbox{\rm [}Co\three\hbox{\rm ]}  & 5627.104  &   3d$^7$\,$^4$F$_e$[9/2] - 3d$^7$ $^2$G$_e$[7/2] & 5.323(-11) &  0.0140 \\
\hbox{\rm [}Co\three\hbox{\rm ]}  & 5888.482   &   3d$^7$\,$^4$F$_e$[9/2] - 3d$^7$ $^2$G$_e$[9/2] & 2.081(-9)   &  0.4001 \\
\hbox{\rm [}Co\three\hbox{\rm ]}  & 5906.783   &   3d$^7$\,$^4$F$_e$[7/2] - 3d$^7$ $^2$G$_e$[7/2] & 7.850(-10) &  0.1500 \\
\hbox{\rm [}Co\three\hbox{\rm ]}  & 6195.455   &   3d$^7$\,$^4$F$_e$[7/2] - 3d$^7$ $^2$G$_e$[9/2] & 8.636(-10) &  0.1200 \\
\hbox{\rm [}Co\three\hbox{\rm ]}  & 6576.309   &   3d$^7$\,$^4$F$_e$[9/2] - 3d$^7$ $^4$P$_e$[5/2] & 1.868(-10) &  0.0480 \\
\hline 
\end{tabular}
\end{table}

A further confirmation of this association is that the observed strength of the
5900\,\AA\ feature typically increases after bolometric maximum. In the context of [Co\three]
emission, this also makes sense since the strength of forbidden lines relative to
other lines should increase as the ejecta density drops.
We show in Fig.~\ref{fig_seq} a montage of spectra for the evolution of SN\,2005cf
and model PDDEL3 \citep{dessart_etal_13yy} from bolometric maximum (i.e., +0\,d) until +80\,d
(we note that model DDC10 studied in detail in D13 does a somewhat better job in the $B$-band 
region, but model PDDEL3 is more compatible with the narrow line profiles of SN\,2005cf; both
models are equally suitable for the present discussion).
As in Fig.~\ref{fig_grid}, we show the flux associated with the [Co\three]\,5888\,\AA\ line
as a filled red area. It is clearly evident that early after bolometric maximum, flux
from that forbidden line contributes to the emergent radiation, and that this flux
contribution increases with time, as observed. We note that at bolometric maximum,
 [Co\three]\,5888\,\AA\ line emission is already a strong coolant of the Co-rich
 layers; [Co\three]\,5888\,\AA\ line photons are not seen earlier on because
 these regions are located below the last scattering/absorbing layer.

There are other [Co\three] lines in the 6000\,\AA\ region (Table~\ref{tab_co3}). Of
all these, the transition at 5888\,\AA\ is the strongest. The transition at 5906\,\AA\ is expected to be
weaker and it also overlaps with the 5888\,\AA\ transition. The transition at 6195\,\AA\ should 
be about 1/3 of the strength of 5888\,\AA\ but the model predicts essentially no flux in this
line. We find that the 6195\,\AA\ suffers absorption by overlapping lines, in particular from
the Si\two\ doublet at 6355\,\AA, which is strong at those epochs. Numerous other lines
are present in this spectral region, while fewer reside in the 5900\,\AA\ region, allowing 
the [Co\three]\,5888\,\AA\ to escape.
We also find that these low-lying states are in LTE with the ground state (they have the same departure coefficients),
although we obtain strong depopulation (and strong departure from LTE) of the ground state 
of [Co\three] through non-thermal ionization and excitation (this departure is also epoch dependent). 
Scattering is thus expected to influence little the formation of these forbidden-line transitions.

\begin{figure}
\epsfig{file=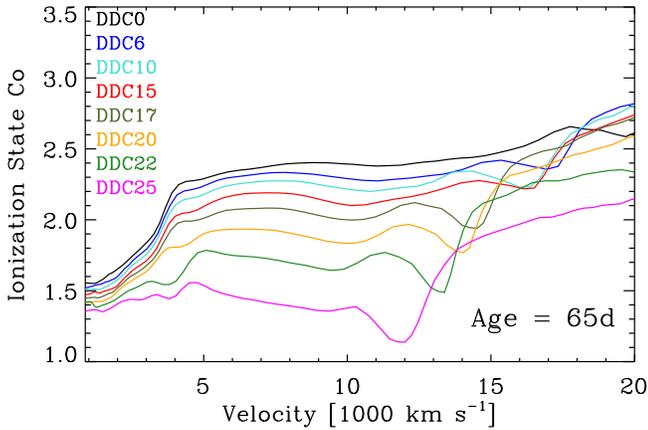,width=8.5cm}
\caption{
Illustration of the Cobalt ionization state in the inner 20000\,\kms\ for our grid
of DDC models at 65\,d after explosion. Models with a higher \nifs\ mass systematically
show a higher Co ionization level, which is directly connected to a range
of spectral colors (bluer for higher \nifs\ mass; Fig.~\ref{fig_grid}).
\label{fig_ion}
}
\end{figure}

We also see that in our simulations, the delayed detonation models systematically show
a range of optical colors at +40\,d, while observations look more similar at this time (Fig.~\ref{fig_grid}). 
This is particularly visible for the low-luminosity SN Ia 1999by,
which shows a stronger [Co\three]\,5888\,\AA\ line and relatively less flux in the red 
than in our model DDC25.
This range in colors in our models reflects the trend in ionization level at the corresponding
epoch, as evidenced by the ionization state of Co in the in the inner ejecta where the spectrum 
forms (Fig.~\ref{fig_ion}). This could arise from problems with the atomic physics, or
from issues with the structure computed for the assumed progenitor model.
Another possible explanation may be that, while the
\nifs\ mass is accurate in each model/observation pair (tightly constrained by the 
peak luminosity), the ejecta mass of 1.38\,\msun, which is kept fixed here, may also vary
and in particular involve sub-Chandrasekhar mass WDs (see, e.g., \citealt{sim_etal_10}).
For a fixed \nifs\ mass, reducing the ejecta mass leads naturally to a significant increase
in ejecta ionization (see, e.g., \citealt{dessart_etal_12c}).

\section{Discussion and conclusions}
\label{sect_conc}

From this work, we unambiguously discard Na\one\,D as the origin of the 5900\,\AA\
feature in SNe Ia after bolometric maximum based on 1) the prediction of [Co\three]\,5888\,\AA\ line 
emission from all our delayed-detonation models after bolometric maximum; 2) the strengthening
of that [Co\three] line with time in the first 2 months after bolometric maximum and 3) the
satisfactory match of our synthetic spectra to the observations.

This finding is not trivial taxonomy because it further
confirms our conclusions from the more theoretical discussion in D13.
Although the [Co\three] line was identified three decades ago in nebular phase spectra of SNe Ia
\citep{axelrod_80} we are discovering the role of [Co\three] 
in SNe Ia {\it as early as bolometric maximum}, and confirming its identification at later times. 
 Lately, a controversial association has been made with Na\one\,D but this association is not 
 supported by models. One explanation for the mis-identification is that [Co\three] lines are not included
in the associated radiative transfer calculations \citep{lentz_etal_01,mazzali_etal_08,tanaka_etal_11},
as was also done in our earlier modeling of SN Ia.
When they are  included, the line shows up \citep{maurer_etal_11}.
Another explanation is that SN radiative transfer is often done by separate codes for ``photospheric"
phase and nebular phase studies, and photospheric codes typically neglect forbidden lines. 
Unfortunately, this creates a boundary between the
two regimes that is artificial. Nebular lines are indeed seen at photospheric epochs, e.g.,
the [Ca\two\,7300\,\AA\ doublet  at the end of the plateau in SNe II-P \citep{dessart_etal_13}
or [Co\three]\,5888\,\AA\ early after bolometric maximum; this work).
Similarly, strong P-Cygni profiles, typical of photospheric-phase spectra persist
well into the nebular phase.

In the future, we will investigate the radiative properties of SNe Ia as they turn nebular. As
shown here, our Chandrasekhar-mass delayed-detonation models exhibit a large range of 
ionization whereas observations appear somewhat degenerate in that respect.
Our simulations differ in \nifs\ mass but have the same ejecta mass of 1.4\,\msun.
While the \nifs\ mass determines the peak luminosity, the ratio of \nifs\ to ejecta mass
is a key ingredient controlling the ionization state of the gas. Hence, we will investigate
whether a range of ejecta masses, tied to a narrow range of \nifs\ to ejecta mass
ratio, can reduce the ionization disparity of our models after maximum.

\section*{Acknowledgments}

LD and SB acknowledge financial support from the European Community through an
International Re-integration Grant, under grant number PIRG04-GA-2008-239184,
and from ``Agence Nationale de la Recherche" grant ANR-2011-Blanc-SIMI-5-6-007-01.
DJH acknowledges support from STScI theory grant HST-AR-12640.01, and NASA theory grant NNX10AC80G.
This work was also supported in part by the National Science Foundation under Grant No. PHYS-1066293 and
benefited from the hospitality of the Aspen Center for Physics.
AK acknowledges the NSF support through the NSF grants AST-0709181 and TG-AST090074.
This work was granted access to the HPC resources of CINES under the
allocation c2013046608 made by GENCI (Grand Equipement
National de Calcul Intensif).

\label{lastpage}

\end{document}